**Fighting seizures with seizures: diffusion and stability in neural systems**

Erik D. Fagerholm[1,*], Chayanin Tangwiriyasakul[2,6,*], Karl J. Friston[3], Inês R. Violante[4], Steven Williams[1], David W. Carmichael[5,6], Suejen Perani[2,7], Federico E. Turkheimer[1], Rosalyn J. Moran[1], Robert Leech[1,§], Mark P. Richardson[2,8,§]

[1] Department of Neuroimaging, King's College London
[2] Department of Basic and Clinical Neuroscience, King's College London
[3] Wellcome Centre for Human Neuroimaging, University College London
[4] School of Psychology, University of Surrey
[5] Developmental Neurosciences, University College London
[6] School of Biomedical Engineering & Imaging Sciences, King's College London
[7] UCL Great Ormond Street Institute of Child Health, University College London
[8] Centre for Epilepsy, King's College Hospital

Corresponding author: erik.fagerholm@kcl.ac.uk

[*,§] These authors contributed equally to this work

**Abstract**

Seizure activity is a ubiquitous and pernicious pathophysiology that, in principle, should yield to mathematical treatments of (neuronal) ensemble dynamics—and therefore interventions on stochastic chaos. A seizure can be characterised as a deviation of neural activity from a stable dynamical regime, i.e. one in which signals fluctuate only within a limited range. *In silico* treatments of neural activity are an important tool for understanding how the brain can achieve stability, as well as how pathology can lead to seizures and potential strategies for mitigating instabilities, e.g. via external stimulation. Here, we demonstrate that the (neuronal) state equation used in Dynamic Causal Modelling generalises to a Fokker-Planck formalism when propagation of neuronal activity along structural connections is considered. Using the Jacobian of this generalised state equation, we show that an initially unstable system can be rendered stable via a reduction in diffusivity (i.e., connectivity that disperses neuronal fluctuations). We show, for neural systems prone to epileptic seizures, that such a reduction can be achieved via external stimulation. Specifically, we show that this stimulation should be applied in such a way as to temporarily mirror epileptic activity in the areas adjoining an affected brain region – thus 'fighting seizures with seizures'. We offer proof of principle using simulations based on functional neuroimaging data collected from patients with idiopathic generalised epilepsy, in which we successfully suppress pathological activity in a distinct sub-network. Our hope is that this technique can form the basis for real-time monitoring and intervention devices that are capable of suppressing or even preventing seizures in a non-invasive manner.

**Introduction**

There is an ongoing interest in treating epilepsy by using brain stimulation[1, 2, 3], as it allows for direct perturbations of the physiological states of neural systems[4, 5]. However, the three basic questions of *when*, *where* and *how* to stimulate for maximum clinical efficacy remain unanswered[6, 7, 8]. Therefore, there is a pressing need for the development of computational frameworks that can be used to model the effect of brain stimulation on neural populations and for the construction of optimal protocols for therapeutic intervention.

Over the last decade, several mathematical models have been proposed to explain the emergence of seizures – primarily with focal epilepsy using electroencephalography (EEG) and electrocorticography (ECoG). For example, Benjamin et al.[9] developed a network-based model to describe a phenomenological model of seizure initiation, while Sinha et al.[10] and Goodfellow et al.[11] developed models to predict neurosurgical outcome. In this study, we propose a model, tested with a combination of EEG and functional magnetic resonance imaging (fMRI) data, in patients with idiopathic generalised epilepsy (IGE).

Dynamic causal modelling (DCM)[12] provides a powerful analytical tool in this setting. DCM was originally designed for inferring latent structure from blood-oxygen-level-dependent (BOLD) time series, by optimising the parameters of a generative model, such as intrinsic connectivity and external driving inputs. In its simplest mathematical form, DCM rests upon a first-order ordinary differential equation (the neuronal state equation), which describes the ways in which neuronal signals change with respect to time. Here, we show that the neuronal state equation can be generalised to account for the ways in which signals change – not only with respect to time – but also with respect to the connectivity architecture of the network within which the signals are constrained to propagate. We show that this additional structural component conforms to a diffusion process and therefore that the generalised neuronal state equation takes the mathematical form of the Fokker-Planck equation. The latter is a partial differential equation that is used to describe the probabilistic evolution of a system, initially studied in the context of Brownian motion[13], and increasingly used in the modelling of neural systems[14, 15, 16]. As we will show, it is the diffusive property of the Fokker-Planck equation that facilitates the suppression of activity via gradient modulation in brain regions prone to epileptic seizures.

This paper comprises three sections. In the first section, we outline the theoretical basis for subsequent applications by showing that the neuronal state equation takes the functional form of the Fokker-Planck equation, when including structural degrees of freedom (connections) within a network. We provide construct validation that the ensuing model provides a more parsimonious explanation of resting state fMRI. Specifically, we show that time series in both epilepsy patients and healthy control subjects are better modelled by the Fokker-Planck equation, compared with the classic (non-structural) neuronal state equation.

In the second section, we show that an initially unstable region can be pushed into a stable regime by virtue of a reduction in network diffusivity. We demonstrate that such a reduction can be achieved by using external stimulation to mirror seizure activity in the area(s) surrounding a pathological brain region.

In the third section, we report a series of simulations incorporating individualised EEG and fMRI data collected in patients with idiopathic generalised epilepsy. We show promising evidence that electrical stimulation is a viable method for suppressing epileptic seizures.

**Methods**

**The generalised neuronal state equation:** Causal models can be considered as the evolution of states $x$ as a static function of themselves. For instance, for the $i^{th}$ region:

$$x_i = f(x, v, \theta), \qquad [1]$$

where $f$ is a nonlinear function, $v$ are external inputs, and $\theta$ are model parameters – usually interpreted in terms of connectivity or rate constants. DCM extends the assumptions in [1] by considering the ways in which states change with respect to time, so that:

$$\frac{dx_i}{dt} = f(x, v, \theta), \qquad [2]$$

which reduces to [1] in the limit that inputs $v$ vary slowly relative to the states $x$.

Extrapolating the logical progression from [1] to [2], we can define a generalised neuronal state equation in which states may vary, not just with respect to time as in [2], but with respect to any arbitrary number of dimensions $\mu$, such that:

$$\sum_{j=1}\frac{dx_i}{d\mu_j} = f(x, v, \theta). \qquad [3]$$

For example, if we consider the simple three-node network in Fig 1:

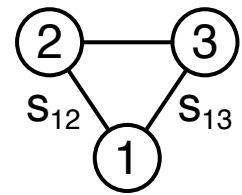

*Figure 1: Three-node network.* The structural connection between nodes 1 and 2 is given by $s_{12}$ and the structural connection between nodes 1 and 3 is given by $s_{13}$.

We describe the evolution of the first node using [3] as follows:

$$\sum_{j=1}^{3}\frac{dx_1}{d\mu_j} = f(x, v, \theta). \qquad [4]$$

We then retain time as the first dimension ($\mu_1 = t$) on the left-hand side and assign the remaining two terms to the rate of change of states along the structural dimensions of the network, such that:

$$\frac{dx_1}{dt} + \sigma \left( \frac{dx_1}{ds_{12}} + \frac{dx_1}{ds_{13}} \right) = f(x, v, \theta), \qquad [5]$$

where $\sigma$ is a (rate) constant of proportionality with dimensions $T^{-1}$; and e.g. $\frac{dx_1}{ds_{12}}$ describes the rate of change of the first node's activity along the edge connecting the first and second nodes. Intuitively, this kind of extension can be thought of as generalising the dynamics of any one point in the brain to a partial differential equation (PDE) that models the spatiotemporal dynamics (e.g., neuronal field models). However, here, the spatial aspect is generalised to edges for connections of a graph or network.

We write the general form of [5] as follows for the $i^{th}$ node in a network of $N$ regions:

$$\frac{dx_i}{dt} + \sigma \sum_{j=1}^{N} k_{ij} \frac{dx_i}{ds_{ij}} = f(x, v, \theta), \qquad [6]$$

where $k_{ij}$ is the adjacency matrix element connecting the $i^{th}$ and $j^{th}$ regions, the inclusion of which ensures that only nearest neighbours in the network are taken into account. Note that in the context of neural systems, the definition of nearest neighbours may also be taken to mean regions that are structurally connected via white matter tracts. This means that any given region or node can have more neighbours than if we were modelling the cortical sheet as a two-dimensional Markov field.

The structural gradients $\frac{dx_i}{ds_{ij}}$ in [6] can be discretized by expressing them in terms of the difference in state values (at a given time) at the $i^{th}$ and $j^{th}$ nodes, such that:

$$\frac{dx_i}{ds_{ij}} \rightarrow x_j - x_i, \qquad [7]$$

which, together with [6], tells us that:

$$\frac{dx_i}{dt} = f(x, v, \theta) + \sigma \sum_{j=1}^{N} k_{ij}(x_i - x_j),  \quad [8]$$

where we see that in the limit of zero structural gradients $(x_i = x_j)$, the second term on the right-hand side vanishes and [8] reduces to the original DCM neuronal state equation in [2]. In other words, every node behaves in the same way and can be described with an ordinary differential equation. In summary, we obtain a third level of neuronal state equations building on the simple causal models in [1], through the classical DCMs in [2], and ending with the generalised form in [8], with each level reducing to the former in the appropriate limiting cases.

**DCM and the Fokker-Planck equation:** The second term on the right-hand side of [8] can be re-written as follows:

$$\sum_{j=1}^{N} k_{ij}(x_i - x_j) = x_i \sum_{j=1}^{N} k_{ij} - \sum_{j=1}^{N} k_{ij} x_j = x_i d_i - \sum_{j=1}^{N} k_{ij} x_j = \sum_{j=1}^{N} (\delta_{ij} d_i - k_{ij}) x_j = \sum_{j=1}^{N} l_{ij} x_j,  \quad [9]$$

where $d_i$ is the degree of the $i^{th}$ node; $\delta_{ij}$ is the Kronecker delta function, which equals unity when $i = j$, and otherwise equals zero; and $l_{ij}$ is the graph Laplacian matrix element connecting the $i^{th}$ and $j^{th}$ nodes, where the graph Laplacian matrix $L$ is defined as the difference between the degree matrix $D$ and adjacency matrix $K$, such that $L = D - K$.

Using [9] we can write [8] as follows:

$$\frac{dx_i}{dt} = f(x, v, \theta) + \sigma \sum_{j=1}^{N} l_{ij} x_j,  \quad [10]$$

and as the graph Laplacian is the discretized version of the Laplace operator[17], the second term on the right-hand side describes a diffusion process, hence lending an interpretation to $\sigma$ as a diffusion coefficient. Therefore, [10] takes the form of the discretized Fokker-Planck equation, with a drift term: $f(x, v, \theta)$ and a diffusion term: $\sigma \sum_{j=1}^{N} l_{ij} x_j$.

**Linear stability analysis:** In order to model neural time series, we assume first order linear interactions, which means [8] can be written as:

$$\frac{dx_i}{dt} = \sum_{j=1}^{N} p_{ij} x_j + \sigma \sum_{j=1}^{N} k_{ij}(x_i - x_j) + \sum_{j=1}^{M} q_{ji} v_j + \omega^{(i)}, \quad [11]$$

where the matrix element $p_{ij}$ reduces to the DCM intrinsic coupling matrix element $a_{ij}$ in the limiting case of zero structural gradients $(x_i = x_j)$; the matrix element $q_{ji}$ reduces to the DCM extrinsic coupling matrix element $c_{ij}$ in the limiting case of zero structural gradients $(x_i = x_j)$; and $\omega^{(i)}$ are non-Markovian fluctuations in the $i^{th}$ region's activity[18]. These fluctuations model deviations from the linear flow of states under an adiabatic approximation. In other words, we assume a centre manifold for the dynamics, which are linear and assign fluctuations tangential to the manifold to $\omega^{(i)}$, which decay rapidly and return to the manifold under the centre manifold theorem[19].

To create a DCM of observable timeseries, we can use [11] as a state space model with fast (analytic) fluctuations $\omega^{(i)}$ and map the latent states $x_i$ to observable quantities with additive observation noise. Equation [11] is used to model all the neural time series presented in this paper, using standard (variational) routines in the Statistical Parametric Mapping (SPM) software.

Extrinsic coupling does not affect resilience to perturbation, as the stability of a linear time invariant (LTI) system is determined by the roots of the characteristic equation $(sI - J)^{-1}$, where the Jacobian $J$ comprises the first two terms of [11] only[20]. Retaining these terms, we can re-write [11] as:

$$\frac{dx_i}{dt} = \sum_{j=1}^{N} p_{ij} x_j + \sigma \sum_{j=1}^{N} k_{ij}(x_i - x_j) = \sum_{j=1}^{N} \left( (p_{ij} - \sigma k_{ij})x_j + \sigma k_{ij} x_i \right), \quad [12]$$

which we can write out explicitly for the three-node network in Figure 1 as follows:

$$\begin{bmatrix} \dot{x}_1 \\ \dot{x}_2 \\ \dot{x}_3 \end{bmatrix} = \begin{bmatrix} p_{11} + \sigma(k_{12} + k_{13}) & p_{12} - \sigma k_{12} & p_{13} - \sigma k_{13} \\ p_{21} - \sigma k_{21} & p_{22} + \sigma(k_{21} + k_{23}) & p_{23} - \sigma k_{23} \\ p_{31} - \sigma k_{31} & p_{32} - \sigma k_{32} & p_{33} + \sigma(k_{31} + k_{32}) \end{bmatrix} \begin{bmatrix} x_1 \\ x_2 \\ x_3 \end{bmatrix}. \qquad [13]$$

The Jacobian of this generalised DCM can thus be written as follows for a network comprising $N$ regions:

$$J = \begin{bmatrix} p_{11} + \sigma \sum_{j \neq 1}^{N} k_{1j} & \cdots & p_{1N} - \sigma k_{1N} \\ \vdots & \ddots & \vdots \\ p_{N1} - \sigma k_{N1} & \cdots & p_{NN} + \sigma \sum_{j \neq N}^{N} k_{Nj} \end{bmatrix}. \qquad [14]$$

**Diffusion and stability:** In 1953, Turing showed that an initially stable dynamical system can be rendered unstable by virtue of a diffusion mechanism, allowing for the emergence of spatial inhomogeneities – now known as Turing patterns[21]. Here, we proceed via similar logic, except we begin with the opposite premise and ask the following question: can we push an initially unstable system (such as a neural system prone to seizures) into a stable regime by altering the system's diffusivity?

To answer this, we multiply the diffusion coefficient $\sigma$ by a constant $\alpha$:

$$\sigma \to \alpha \sigma. \qquad [15]$$

By multiplying $\sigma$ by some unknown quantity $\alpha$ in this way we can determine – in the subsequent linear stability analysis – whether this change increases or decreases diffusivity in order to render an initially unstable system stable.

We assume that the system described by [14] is initially unstable, such that the sum of the Real components of its eigenvalues (given by the trace) is positive:

$$trJ = p_{11} + \sigma \sum_{j \neq 1}^{N} k_{1j} + \cdots + p_{NN} + \sigma \sum_{j \neq N}^{N} k_{Nj} > 0, \qquad [16]$$

We then transform [14] via [15] to obtain:

$$J' = \begin{bmatrix} p_{11} + \alpha\sigma \sum_{j \neq 1}^{N} k_{1j} & \cdots & p_{1N} - \alpha\sigma k_{1N} \\ \vdots & \ddots & \vdots \\ p_{N1} - \sigma k_{N1} & \cdots & p_{NN} + \sigma \sum_{j \neq N}^{N} k_{Nj} \end{bmatrix}, \qquad [17]$$

which we assume has been rendered stable due to the altered diffusion coefficient, such that the Real component of the sum of its eigenvalues is now negative:

$$trJ' = p_{11} + \alpha\sigma \sum_{j \neq 1}^{N} k_{1j} + \cdots + p_{NN} + \sigma \sum_{j \neq N}^{N} k_{Nj} < 0, \qquad [18]$$

which, together with [16], means that:

$$trJ' = trJ + \sigma(\alpha - 1) \sum_{j \neq 1}^{N} k_{1j} < 0. \qquad [19]$$

We then know from [16] that $trJ > 0$. Furthermore, we know that diffusion coefficients are necessarily positive, i.e. $\sigma > 0$, given that they play the role of rate constants[22]. Therefore, the only way in which [19] can be satisfied is if $\alpha$ is less than unity:

$$(\alpha - 1) \sum_{j \neq 1}^{N} k_{1j} < 0 \implies \alpha < 1, \qquad [20]$$

i.e. an initially unstable system can be rendered stable by virtue of a reduction in the diffusion coefficient.

**Diffusivity can be altered via external driving inputs:** In practice it is not possible to change the diffusion coefficient as in [15], due to the fact that it is an intrinsic property of the dynamical system described by [14]. However, as the diffusion coefficient quantifies diffusivity via Fick's law[23], we recover an important piece of information from [20]; namely, that if we want to push a system toward stability, we must act so as to decrease diffusivity.

If we look at the governing equation of motion [11], we note that the diffusion term $\sigma \sum_{j=1}^{N} k_{ij}(x_i - x_j)$ comprises three factors: 1) the diffusion coefficient $\sigma$, which we noted above cannot be changed; 2) the adjacency matrix element $k_{ij}$ which, similar to $\sigma$, is intrinsic to the system and is therefore also unchangeable (without resorting to surgical intervention); and finally 3) the gradient $(x_i - x_j)$. It is this last factor that we can influence by applying specific external driving inputs in strategic locations, in order to decrease gradients and thus to decrease diffusivity. For instance, let us consider the same three-node system shown in Figure 1 and assume that the bottom node is prone to instabilities (Fig. 2A).

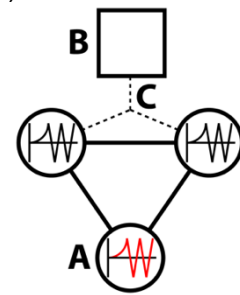

*Figure 2: Gradient reduction via stimulation. A) The seizure-prone node. This displays the activity shown in the red graph therein. B) External driving input is applied in such a way as to allow the two neighbouring nodes to mirror the activity of the seizure-prone node in A). C) Extrinsic coupling to the two regions neighbouring A).*

In the case of epilepsy, we therefore propose a form of intervention in which we apply external driving input (Fig. 2B) in such a way as to allow the two neighbouring nodes (Fig. 2C) to mirror the activity in the region that is prone to seizures.

In summary, we have derived a generalised dynamic causal model of neuronal activity that can generate empirical timeseries. In what follows, estimate the parameters of the DCM using empirical timeseries from patients with epilepsy. Equipped with these parameters, we can then evaluate the Jacobian and simulate the effects of an intervention that moves the ensemble or population dynamics implicit in [11] from a regime of instability (i.e., seizure activity) to a one of stability.

**The seizure network**: In the analyses below, we use a network comprising the following regions: frontal mid, frontal mid orbital, precuneus, and thalamus. This network is known to play an important role in generating generalised spike and wave (GSW) discharges[24] and, for simplicity, we will refer to it henceforth as the 'seizure network'.

**Participants and data acquisition**: We analyse data recorded from 15 patients (6 male) with juvenile myoclonic epilepsy (JME) with a mean age of 24.5 years, and 15 age-matched healthy controls (5 male) with a mean age of 25.2 years (see Supplementary Table I). The data were acquired at the Institute of Psychiatry Psychology and Neuroscience (IoPPN), King's College London. The patients did not have any neurological diagnoses other than epilepsy and had no history of drug or alcohol misuse. The study was approved by the Riverside Research Ethics Committee (12/LO/2005) and all participants signed a written informed consent form prior to the study, according to the declaration of Helsinki (2013). All participants underwent a resting state simultaneous EEG-fMRI and a diffusion tensor imaging (DTI) session with 32 directions and b = 1500 s/mm². A 3T scanner (MR750, GE Healthcare) was used to acquire 300 echo-planer images (3.3 × 3.3 × 3.3 mm, field of view 211 mm, repetition time 2.16 s, echo time 25 ms, flip angle 75°, 36 slices, slice thickness 2.5 mm).

**Bayesian model inversion:** The stochastic differential equation in [11] furnishes a dynamic causal model, where fast fluctuations are assumed to be small. This means that, assuming that the underlying dynamics can be modelled by [11], the model parameters $\theta = (p, \sigma)$ can be recovered from observations of BOLD signals (Fig. 3A).

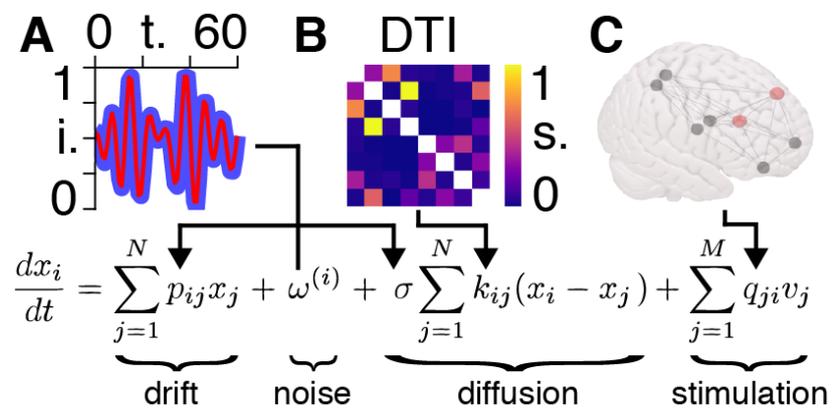

*Figure 3: Equation of motion.* ***A)*** *BOLD signal intensity (i.) for an example time course (in seconds) in one patient (blue), together with the time series estimate (red) following Bayesian model inversion. This inversion provides posterior densities over the matrix elements $p$ in the drift term, and the diffusion coefficient $\sigma$ in the diffusion term, informed by a non-Markovian noise process $\omega$.* ***B)*** *Example of an adjacency matrix from DTI data, from which we obtain matrix elements $k$ in the diffusion term.* ***C)*** *The eight regions of the seizure network, in which the left and right frontal mid regions (red) give rise to unstable activity. Stimulation would then be applied to one or more of the remaining (black) regions in a way that mirrors the activity of the red nodes.*

$$\frac{dx_i}{dt} = \underbrace{\sum_{j=1}^{N} p_{ij} x_j}_{\text{drift}} + \underbrace{\omega^{(i)}}_{\text{noise}} + \underbrace{\sigma \sum_{j=1}^{N} k_{ij}(x_i - x_j)}_{\text{diffusion}} + \underbrace{\sum_{j=1}^{M} q_{ji} v_j}_{\text{stimulation}}$$

As the parameter space associated with the eight-region seizure network is larger than can be accommodated by the time points available for each scan, we reduce the number of free parameters via functional connectivity priors[25, 26]. This allows us to constrain the optimization such that we retain ~10× as many time points as free parameters. The adjacency matrix elements $k_{ij}$ are not included as free parameters as they are known *a priori* from diffusion imaging (Fig. 3B). We use generalised or variational Bayesian filtering (specifically, Dynamic Expectation Maximisation (DEM))[27] to a) infer the latent states; b) estimate the parameters; and c) hyperparameters, i.e. the precision components of fluctuations on the states and observation noise. After recovering the posterior densities, hyperparameters, and variational free energies on an individual level for all 30 subjects, we then obtain two group-level models by performing Bayesian model averaging separately across 15 patients and 15 controls. These averaged models are used as the bases for all the forward generative models using external stimulation as a means of reducing diffusivity (Fig. 3C).

**Data preprocessing**: To pre-process the fMRI data, we use the statistical parametric mapping (SPM8, r613) software running on MATLAB (R2016b), together with the FIACH[28] package for R (3.2.2). First, we convert the data from DICOM to Nifti formats. We then delete the first four volumes of each session to avoid magnetic saturation effects. We subsequently re-align all images to the first remaining volume. To correct for possible artefacts, we apply the FIACH toolbox to the BOLD time series. We then normalise all data into standard MNI space with 2 mm isotropic voxels. All images are then spatially smoothed using a Gaussian filter of 8 mm fullwidth at half maximum. The BOLD signal is filtered between 0.04 – 0.07 Hz[29]. This frequency range was chosen to minimise the overlap between the BOLD signal and possible breathing and pulsation artefacts. We parcellate the brain into standard 90 automated anatomical labelling (AAL) regions (excluding the cerebellum)[30]. Finally, we apply principal component analysis (PCA) to the voxel time series within each region, from which we retain the first principal component to summarise the activity in each region[31]. Probabilistic tractography is used to pre-process DTI data using the iFOD2 algorithm within the MRtrix software[32]. Streamlines are filtered using SIFT[33], resulting in $10^7$ streamlines. We estimate a $90 \times 90$ structural connectivity matrix according to the number of streamlines connecting each pair of regions, normalised by their combined volumes. To reduce inter-subject variability, we then normalise each structural connectivity matrix relative to its maximum value. For each subject, we binarize the structural connectivity matrix according to 18 thresholds between 10% and 95% (in steps of 5%). Equipped with the structural measures $k_{ij}$ and the summaries of regional activity in our seizure network, we estimate the latent states $x_i$ and parameters of the DCM – crucially including the diffusion coefficient $\sigma$ (see equation [11]). Our special interest here lies in the diffusion coefficient and its role in mediating dynamical instability that we associate with seizure activity.

**Results**

**Diffusivity is higher in the patient group:** We find that the patient group has higher diffusivity as compared with the control group across structural adjacency matrix thresholds used to define $k_{ij}$ (see Figure 4). The higher diffusivity in the patient group across all thresholds speaks to the approach of suppressing seizure activity by decreasing diffusivity.

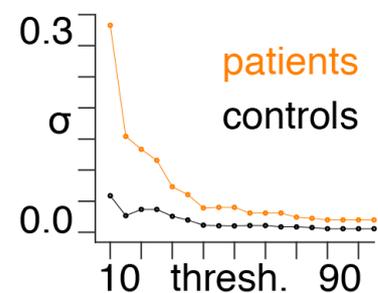

*Figure 4: Diffusion coefficients.* The diffusion coefficient ($\sigma$) as a function of structural (DTI) adjacency matrix threshold (%) for patients and controls following Bayesian model averaging.

**Stability is lower in the patient group:** We find that the patient group is associated with lower levels of stability – as quantified with the sum of the Real components of the eigenvalues

– as compared with the control group across structural adjacency matrix thresholds (see Figure 5).

The lower stability in the patient group speaks to the approach of aiming to increase stability. All subsequent results shown in this paper are calculated using a threshold of 50% to define structural connectivity (i.e., $k_{ij}$).

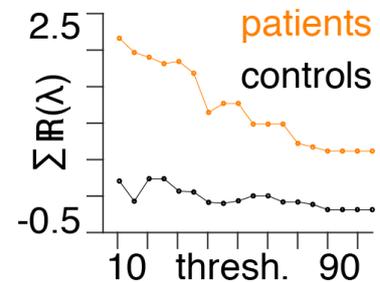

*Figure 5: Stability.* The sum of the Real components of the eigenvalues of the Jacobian ($\sum \mathbb{R}(\lambda)$) – measuring intrinsic stability of neural dynamics – as a function of structural (DTI) adjacency matrix threshold (%) for patients and controls following Bayesian model averaging.

**Accounting for diffusion improves models:** In order to determine whether we are licensed to include the diffusion term in the equation of motion [11], we first optimise the full model including a non-zero diffusion coefficient $\sigma$ (i.e., including diffusion) and subsequently use Bayesian model reduction[34, 35] to estimate the evidence for the reduced model in which $\sigma = 0$ (i.e. excluding diffusion). We specify the reduced model by setting the prior variance over the $\sigma$ parameter to zero, where $\sigma$ is given a prior mean of zero. We show, in both patients and controls, that the variational free energies and associated probabilities are higher for the full models including diffusion, than in the reduced models excluding diffusion (Figure 6). This Bayesian model comparison provides evidence for our assumption that the diffusion term in the equation of motion [11] is necessary to explain these BOLD time series.

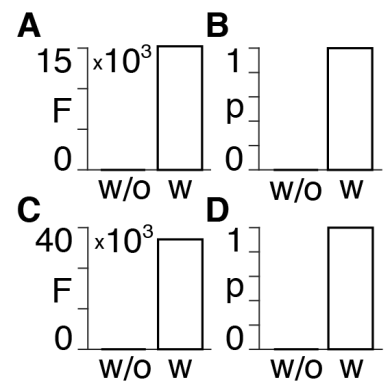

*Figure 6: Bayesian model reduction.* **A)** *Control group. Approximate lower bound on log model evidence afforded by the free energy (F) following Bayesian model reduction for the reduced model without (w/o) diffusion and the full model with (w) diffusion.* **B)** *Probabilities derived from the log evidence in A).* **C) & D)** *Same layout as A) & B), but for the patient group.*

**Ictal onset perturbation:** Following Bayesian model inversion and model averaging in the patient group, we use the resulting parameters to create an *in silico* seizure network, in which we can perturb the mid-frontal sources with ictal onset activity taken from EEG measurements (Figure 7A & B). We then obtain the response of the mid-frontal regions to this exogenous stimulation, which we see climbs in an uncontrolled manner, due to the associated positive Real eigenvalue (Figure 7C).

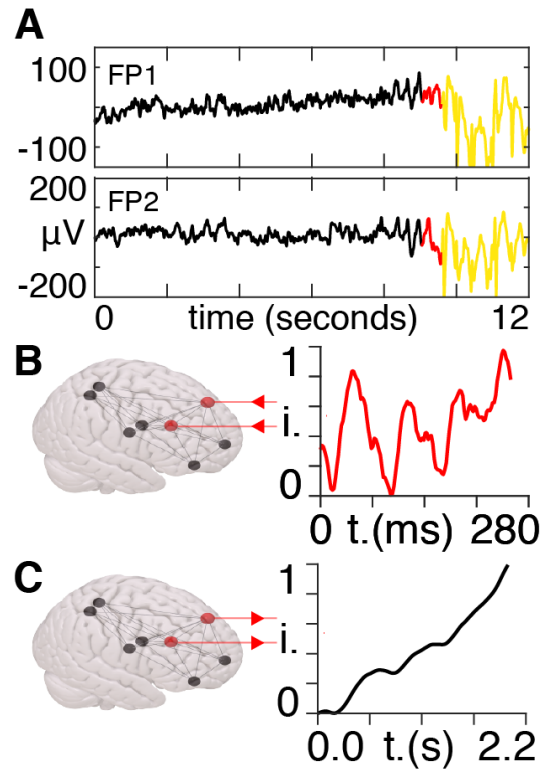

*Figure 7:* **Ictal onset perturbation.** *A) Normalised EEG activity from frontal lobe (FP1 & FP2) activity. Ictal activity is shown by the yellow section and ictal onset activity is shown by the red section. B) The seizure network shown in MNI space (left) with the mid-frontal region (left & right) indicated by the red nodes. The mean ictal onset activity (right), corresponding to the red sections in A), is used as the external driving input and is supplied to the red nodes, as indicated by the inward-pointing red arrows. C) The mean response of the mid-frontal region to the stimulus in B), as indicated by the outward-pointing red arrows. Note that the time scale is longer due to the model inversion having been applied to BOLD data.*

**Seizure suppression:** Using the same setup as in Figure 7, we again run the forward model. However, this time – in addition to the ictal driving stimulus – we supply additional external stimulation to all nodes *except* the mid-frontal region, in a way that mirrors the ictal onset activity in Figure 7B. We show that, in agreement with our theoretical predictions, reducing activity gradients in this way results in the response of the unstable mid-frontal region being suppressed (Figure 8).

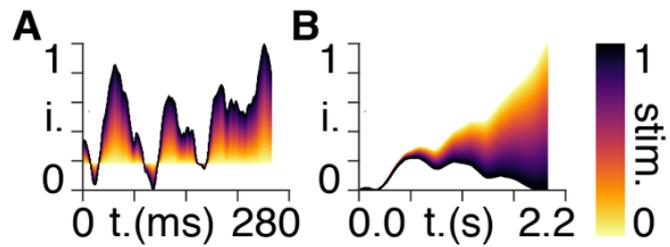

*Figure 8: Simulating seizure suppression. **A)** External stimulation applied to all nodes except for the mid frontal region for 1000 forward models ranging from zero stimulation (yellow) to a stimulation profile that perfectly mirrors the ictal onset activity in Figure 7B (black). **B)** Response of the mid frontal region for the same 1000 forward models in A) with matching colours, i.e. the yellow response corresponds to zero stimulation and the black response corresponds to a stimulation profile that perfectly mirrors the ictal onset activity in Figure 7B.*

Note that although these simulations are based on fMRI data collected in patients with epilepsy, these patients did not experience seizures while in the scanner: the ictal onset activity collected with EEG was recorded separately. The output of the model in Figure 7C and Figure 8B should therefore not be viewed as mimicking a seizure-like BOLD response. Instead, what we show here is that, in response to a perturbation in the form of real ictal onset activity (Figure 7B and Figure 8A), an initially unstable system with a diverging response can be suppressed – a result that we suggest could be clinically advantagous in the treatment of epilepsy.

**Minimum effective stimulation**: In the previous section we stimulated all nodes except for the mid-frontal region (which receives the driving input) to demonstrate proof of principle. However, for treatment purposes, it is clearly better to stimulate as few regions in the seizure network as possible, in order to remain minimally invasive. We therefore proceed by using each of the nodes of the seizure network individually in turn as the stimulation target. This allows us to determine the order of the regions, ranked in terms of the lowest stimulation strength required to suppress the mid-frontal response (Table 1).

*Table 1: Ranking order in terms of the stimulation strength required to suppress the BOLD response in the mid-frontal region to the same extent as in Figure 8B by stimulating a single region in the seizure network. Stimulation strengths are presented relative to the lowest value (thalamus left), which is assigned a value of unity. N/A values are assigned if it is not possible to suppress the BOLD response in the mid-frontal region by targeting the corresponding single region.*

| Rank | Region | Stimulation |
|---|---|---|
| 1 | Thalamus (left) | 1 |
| 2 | Frontal Mid Orbital (right) | 20 |
| 3 | Precuneus (left) | 76 |
| 4 | Precuneus (right) | 82 |
| 5 | Frontal Mid Orbital (left) | N/A |
| 6 | Thalamus (right) | N/A |

Note that the mid-frontal region is not included in this ranking as it is being supplied with the driving input in the form of ictal onset activity and our technique requires mirroring this activity in neighbouring nodes in the network, in order to decrease gradients and thus diffusivity.

Using this *in silico* stimulation protocol, we find that the thalamus (left) requires the lowest stimulation strength (by a considerable margin) to suppress activity in the unstable mid-frontal region relative to the other six regions in the seizure network.

**Discussion**

We began by showing that the neuronal state equation from Dynamic Causal Modelling takes the form of the Fokker-Planck equation when generalised to account for the propagation of neuronal activity over structured connections. As the Fokker-Planck equation entails a conservation of probability[36], the generalised state equation implies a conservation of neuronal activity in the brain – which can be plausibly motivated in terms of conservative aspects of neuronal message passing, such as the balance between excitation and inhibition[37, 38]. This balance lies at the base of functional modes, which are found (particularly in the cortices) to repeat across scales in the brain[39]. The canonical computational units at the most elementary scale take on various ratios of excitatory and inhibitory neurons. However, the current consensus is that the basic unit of the cortical system is the pyramidal interneuron gamma network (PING)[40]. The PING configuration is made up of a pyramidal excitatory neuron (PN) and a fast spiking inhibitory parvalbumin interneuron (IN). One can interpret the diffusion term in equation [11] as an intrinsic mode of interaction between such neuronal units across spatial scales, where the responsible mechanism could be due to: 1) neurotransmission at the microscopic scale of individual neurons, 2) electrical impulses at the mesoscopic scale of microcircuits, and 3) long-range white matter connections between brain regions at a macroscopic scale, as captured by our DTI data.

The intervention technique we are proposing is a departure from the status quo, which usually involves the opposite approach – namely, increasing inhibition[41, 42, 43]. Our method was shown to work in the context of forward generative models, parameters of which were optimised from data collected in epilepsy patients. Specifically, we applied stimulation in such a way as to decrease activity gradients between unstable regions and their neighbouring nodes, thereby decreasing diffusivity.

It is important to emphasize that phasic stimulation cannot change the system's long-term stability, as this is determined by intrinsic properties, such as the diffusion coefficient and connectivity. Instead, we are proposing that it should be possible to temporarily suppress pathological regional activity – achieving a 'quasi-stability' – sufficiently long enough to attenuate seizure activity. Furthermore, one need not wait for a seizure to be in progress in order to activate stimulation. Rather, it may be advantageous to continuously minimize activity gradients in the areas surrounding a known pathological region (e.g., epileptogenic zone), thereby preventing seizures from occurring in the first place.

Overall, the approach proposed here provides a novel framework to address the three fundamental questions of *when*, *where,* and *how* to stimulate the brain in order to suppress pathological activity in the context of epilepsy. In particular, we investigated the impact of stimulation strength in relation to the number of stimulation sites and proposed specific stimulation timings and profiles, in such a way as to achieve modulation of activity gradients. As our technique relies upon supplying otherwise healthy regions of the brain with stimulation that mirrors pathological activity, it is clearly clinically advantageous to target as few regions as possible. It is for this reason that we focused on targeting a single region and found that the thalamus required the lowest stimulation strength, in line with the broad literature showing the thalamus to be a key region in seizure generation[44, 45, 46, 47]. However, it is in principle possible to target multiple nodes. The trade-off between the number of target nodes and stimulation strengths required is to be determined going forward in practical applications of this technique on a patient-by-patient basis – informed by the specific pathology of the individual.

Stimulation techniques are currently available using both invasive and non-invasive methods. The fast computational processing times associated with our strategy render it compatible with closed-loop approaches, which are increasingly seen as providing the greatest clinical efficacy in delivering personalised therapy[48]. Furthermore, the methodology presented is not limited to applications in epilepsy. For instance, it may be beneficial to use the same gradient reduction technique in the treatment of e.g. disorders associated with cortical spreading depression (CSD)[49].

There are similarities between our approach and coordinated reset strategies, given that our results support targeting several stimulation sites in a spatially and temporally coordinated manner[50, 51]. However, a critical difference in our study is that we demonstrate that abnormal activity can be mitigated by changes in network diffusivity via the counter-intuitive approach of *increasing* excitation in a strategic manner, rather than by following a phase-resetting mechanism. Specifically, we demonstrated that unstable activity can be suppressed by modulating the neighbourhood of affected brain regions, with the stimulation profile and timing chosen in such a way as to mirror the pathological activity – hence 'fighting seizures with seizures'.

***Supplementary Table I:***
*Demographics of the subjects in this study*

| Healthy controls | | | Patients | | |
|---|---|---|---|---|---|
| Subject | Age | Gender | Subject | Age | Gender |
| H01 | 24 | M | P01 | 16 | F |
| H02 | 22 | M | P02 | 15 | F |
| H03 | 24 | F | P03 | 17 | F |
| H04 | 20 | F | P04 | 26 | F |
| H05 | 26 | M | P05 | 16 | F |
| H06 | 25 | F | P06 | 20 | M |
| H07 | 23 | M | P07 | 25 | M |
| H08 | 23 | F | P08 | 37 | F |
| H09 | 22 | F | P09 | 28 | F |
| H10 | 20 | F | P10 | 30 | F |
| H11 | 28 | F | P11 | 35 | M |
| H12 | 17 | F | P12 | 21 | M |
| H13 | 44 | F | P13 | 22 | F |
| H14 | 39 | F | P14 | 39 | M |
| H15 | 21 | M | P15 | 20 | M |

**Code availability:** We make all code used to produce the results in this paper available in the following public repository: https://github.com/allavailablepubliccode/Diffusion

**Author contributions:** S.P. collected the data under the supervision of M.P.R. and D.C.; C.T. preprocessed the data; all authors designed and performed research, analysed data, and wrote the paper.

**Acknowledgements:** We thank all participants and patients for generous involvement in this study, as well as the radiography staff at King's College London for their excellent technical support in data acquisition. The study was performed at the NIHR-Wellcome Trust King's Clinical Research Facility at King's College Hospital. E.D.F. and R.L were funded by the MRC (Ref: MR/R005370/1); I.R.V. was funded by the Wellcome Trust (Ref: 103045/Z/13/Z) and the BBSRC (Ref: BB/S008314/1); K.J.F. was funded by a Wellcome Principal Research Fellowship (Ref: 088130/Z/09/Z); R.J.M was funded by the Wellcome/EPSRC Centre for Medical Engineering (Ref: WT 203148/Z/16/Z). M.P.R. is supported in part by the National Institute for Health Research (NIHR) Biomedical Research Centre at the South London and Maudsley Hospital NHS Foundation Trust and King's College London, as well as the Engineering and Physical Sciences Research Council (EPSRC) Centre in Predictive Modelling in Healthcare grant number EP/N014391/1, and by the MRC Centre for Neurodevelopmental Disorders grant number MR/N026063/1. This study was also supported by a Medical Research Council (MRC) Programme Grant number MR/K013998/1 and a Wellcome/EPSRC Centre for Medical Engineering Programme Grant number WT/203148/Z/16/Z. The authors would also like to acknowledge support from the Data to Early Diagnosis and Precision Medicine Industrial Strategy Challenge Fund, UK Research and Innovation (UKRI), the National Institute for Health Research (NIHR), the Biomedical Research Centre at South London, the Maudsley NHS Foundation Trust, and King's College London.

**Competing interests:** The authors declare no competing interests.